# Direct optical sensing of single unlabeled small proteins and super-resolution microscopy of their binding sites

Marek Piliarik and Vahid Sandoghdar[*]

*Max Planck Institute for the Science of Light and Friedrich Alexander University Erlangen-Nuremberg, 91058 Erlangen, Germany*

**Abstract**

More than twenty years ago, scientists succeeded in pushing the limits of optical detection to single molecules using fluorescence. This breakthrough has revolutionized biophysical measurements, but restrictions in photophysics and labeling protocols have motivated many efforts to achieve fluorescence-free single-molecule sensitivity in biological studies. Although several interesting mechanisms using vibrational spectroscopy, photothermal detection, plasmonics or microcavities have been proposed for biosensing at the single-protein level, no method has succeeded in direct label-free detection of single proteins. Here, we present the first results using interferometric detection of scattering (iSCAT) from single proteins without the need for any label, optical nanostructure or microcavity. Furthermore, we demonstrate super-resolution imaging of protein binding with nanometer localization precision. The ease of iSCAT instrumentation promises a breakthrough for industrial usage as well as fundamental laboratory experiments.

Highly sensitive detection of biological entities is of central importance not only for laboratory and clinical research but also for public health, environmental monitoring and pharmaceutical industry[1]. Various strategies based on mechanical[2], electrochemical[3] and optical[4,5] interactions have been pursued for different applications. A particularly interesting approach is based on measuring the refractive index change that analyte molecules incur upon binding. To achieve this, researchers have exploited resonance shifts in surface plasmons[6], microresonators[7,8] or nanoparticle plasmons[9,10,11]. Each method relies on field confinement in a structure that acts as a large optical label with characteristic resonance. These techniques confront fundamental challenges posed by several issues: 1) increasing the sensitivity towards single molecules comes at the cost of reducing the size of the sensor active area or "hotspot", 2) a clear digital detection is difficult to achieve because the sensor sensitivity follows a smooth functional form within the hotspot, and 3) the spatial coordinates of the analyte are not accessible. Nevertheless, devices based on surface plasmons play a major role in the commercial market[12]. Here, we demonstrate that direct interferometric detection of scattering[13,14] (iSCAT) offers a large-area label-free sensing methodology able to count single proteins with molecular weight less than 60 kDa.

The most commonly used optical detection in biomedical laboratories is based on fluorescence[15]. However, the finite size of the marker, the extra complexity of labeling and the severe difficulties caused by photobleaching have motivated many groups to pursue strategies for label-free and absorption-free detection of biological species. One interesting alternative is to exploit nonlinear spectroscopy of the vibrational levels[16]. This method has an exquisite selectivity but its sensitivity falls short of single-molecule detection. Another popular approach relies on linear optics and the detection of the refractive index change due to analyte binding. The oldest and commercially available implementation of this technique records the shift of the plasmon resonance of a gold-coated prism[17], whereby the specificity and selectivity are provided by surface chemistry. While submonolayer detection is readily achievable in such a device, a single molecule does not manifest a significant change in the refractive index of the sensor medium. To increase the sensitivity, confined modes of plasmonic nanostructures and dielectric microresonators have been investigated, but these techniques are intrinsically limited. First, there is a compromise between the size of the sensor active area and its sensitivity. Higher sensitivity comes through confined intensity distributions and, thus, at the cost of more restricted active area, fewer binding receptors, and the requirement for thinner functionalization[18]. Second, the strong confinement in hotspots is accompanied with a large gradient of sensor response over its active area. As a result, a clear "yes-no" detection is not possible. Furthermore, no spatial information is available about the location of the individual proteins. These shortcomings make it very difficult to characterize the performance of a sensor in a quantitative and robust fashion[19,20]. Indeed, the recent reports of single-protein sensitivity have relied on consistency arguments and comparisons with theoretical simulations [8,11,10].

In this work we report on the direct label-free detection and imaging of individual proteins via the interference of the light created by Rayleigh scattering and the reflection of the incident laser beam[13,21,22]. Measurement of the light scattered by a single protein in its natural environment and far from its resonant absorption confronts two main challenges. First, the small signal of a single biomolecule has to be larger than the noise of the detector. Second, the signal has to be distinguished over fluctuations originating from other scattering sources within the optical path. In the past, scattering and interferometric methods have provided measurements of the absolute molar mass and concentration in bulk samples[23] and of protein submonolayers[5]. However, the extremely small scattering cross sections of proteins have hampered the extension of these methods to the single-molecule level.

Fig. 1 displays the essence of our experimental setup. A laser beam illuminates a glass substrate, and its partial reflection at the substrate-water interface is used as the reference for a homodyne interferometric detection[13]. Molecules adsorbed on the substrate and any optical surface inhomogeneities generate scattering, which is collected by the microscope objective (see inset in Fig. 1a). The reference and scattered components reach a CMOS camera as planar and converging spherical waves, respectively. Because the two optical fields are coherent, they interfere and result in the detected power ($P_{det}$) given by

(1) $$P_{det} = P_{inc}(r^2 + s^2 + 2rs \cos \varphi) = P_{ref} + P_{scat} + P_{int} \;.$$

Here, $P_{inc}$ is the incident power, $r$ is the field reflectivity of the glass-water interface, $\varphi$ denotes a phase (mainly determined by the Gouy phase shift[13]), and the unitless parameter $s$ is related to the particle scattering cross section and thus polarizability. While the second term ($P_{scat}$) describes the scattering power of the object, the third term ($P_{int}$) represents the beating of the reference (local oscillator) and scattered fields.

The strength of Rayleigh scattering by a subwavelength nano-object is determined by the incident electric field and its polarizability ($\alpha$), or equivalently cross section ($\sigma$). Textbook formulae $\alpha = 3V(n_s^2 - n_m^2)(n_s^2 + 2n_m^2)^{-1}$ and $\sigma = \frac{8}{3}\pi^3 \alpha^2 (\lambda/n_m)^{-4}$ provide estimates of these quantities, where *V* is the object volume, $n_s$ denotes its refractive index, $n_m$ stands for the refractive index of the surrounding medium, and $\lambda$ is the illumination wavelength. For a small biomolecule such as albumin with molecular weight of 60 kDa, effective scattering radius of 3.7 nm[24] and refractive index of 1.44, one obtains $\sigma = 10^{-11}$ μ$m^2$ at $\lambda = 405$ nm in an aqueous environment. It follows that an incident power of $P_{inc}$ (in units of photons per second) yields $(\sigma/A)P_{inc}$ scattered photons per second, where we take the characteristic area *A* associated with the detection of a single protein to be the area of a diffraction-limited spot (DLS). Assuming ideal collection efficiency, no detection losses and $A = \pi(100\ nm)^2$, the power registered on the detector is $P_{scat} = (3 \times 10^{-10})P_{inc}$. This is to be compared with the power of the reference beam $P_{ref} = (6 \times 10^{-3})P_{inc}$ for $r^2 = 6 \times 10^{-3}$ at the water-glass interface. The divide of more than $10^7$ between $P_{scat}$ and $P_{ref}$ puts a severe limitation on the dynamic range of a real detector and renders their simultaneous measurement very difficult. However, the third term in Eq. (1) corresponding to $P_{int} = 2\sqrt{P_{scat}}\sqrt{P_{ref}}$ remains as large as $P_{int} = (3 \times 10^{-6})P_{inc} = (5 \times 10^{-4})P_{ref}$. Hence, we express the iSCAT signal originating from a nano-object in terms of the contrast $\frac{P_{int}}{P_{ref}}$.

The visibility of $P_{int}$ over $P_{ref}$ depends on the noise level of the latter. In the optimal situation, where the intensity fluctuations are dictated by the photon shot noise the condition for deciphering $P_{int}$ within integration time $\tau$ becomes $P_{int}\tau > \sqrt{P_{ref}\tau}$. Therefore, detection of a small protein in $\tau = 100\ ms$ should be possible for $P_{inc} > 10^{10}$ photons s$^{-1}$ DLS$^{-1}$, corresponding to 5 nW focused to a DLS. In a realistic laboratory experiment, losses in the collection of scattered light, through the optical elements, and in the detector call for a larger power. The actual parameters and considerations of our experiment can be found in the Supplementary Information. Here, it suffices to note that we performed our measurements at a rate of 3000 frames per second under $P_{inc} = 15$ μW per DLS. We also emphasize that the corresponding intensity is many orders of magnitude away from the damage threshold of biological matter.

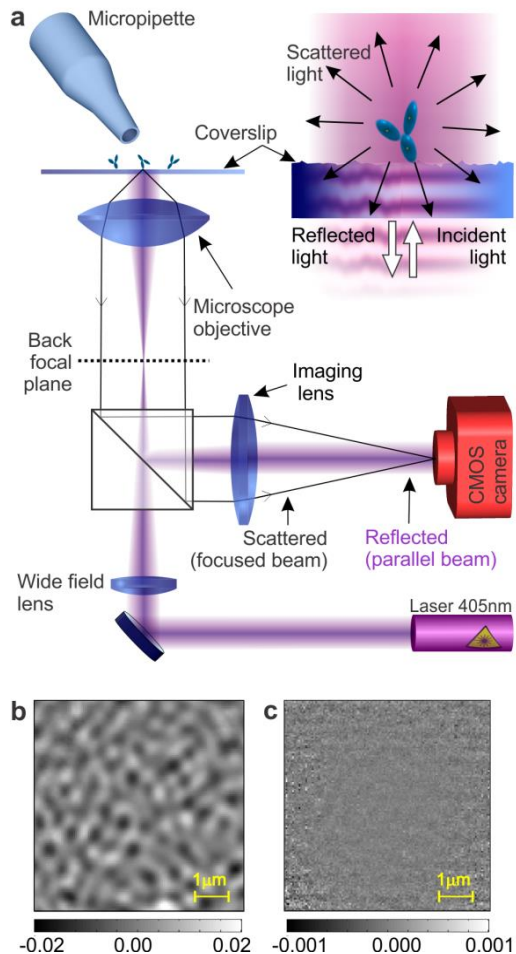

**Fig. 1 Experimental configuration of the interferometric scattering biosensor. a,** Laser light is focused on the back focal plane of a microscope objective. The focal plane of the imaging lens coincides with the back focal plane of the objective. A pulled capillary is used as a micropipette to deliver the analyte locally to the field of view of the microscope. The inset illustrates a zoom of the interactions of the incident, reflected, and scattered light waves. **b,** A typical interferometric image of the bare sensor surface. **c**, The differential image of the sensor surface (see text for details).

Fig. 1b shows a typical iSCAT camera image of a naked substrate. Surface corrugations of the glass coverslip and possibly small local variations in the refractive index result in contrast fluctuations at the level of $4\times10^{-2}$, which is considerably larger than the signal expected from a single protein. However, because the background associated with the surface roughness is static, we can mask it by subtracting consecutive images. Fig. 1c displays the resulting differential image. It follows that subtracting images recorded prior and after the arrival of analyte molecules allows us to distinguish the latter on a background that is only limited by shot noise.

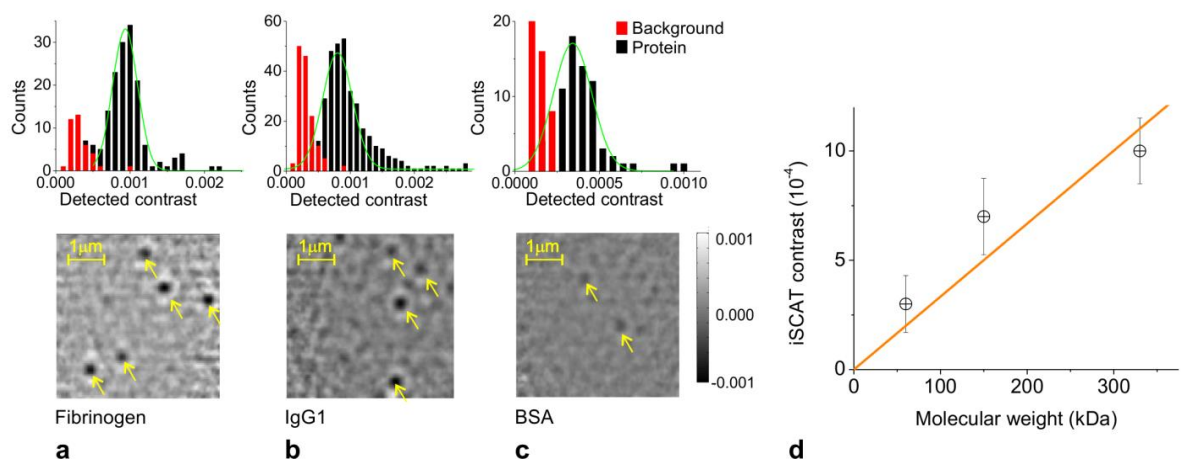

**Fig. 2 iSCAT images and histograms for three different proteins. a, b, c,** Examples of differential images of fibrinogen, IgG1, and BSA, respectively. Individual molecules are marked with arrows. A histogram of the contrasts measured prior (background) and during protein detection is shown above each image. **d**, Plot of the iSCAT contrast (histogram peak) as a function of the molecular weight of the proteins. Error bars indicate the FWHM of corresponding histograms.

Fig. 2a-c presents examples of differential images after solutions of fibrinogen (a), mouse immunoglobulin (IgG1) (b) and bovine serum albumin (BSA) (c) were introduced to the vicinity of the a surface activated by N-hydroxysuccinimide (NHS). The images were created using averages of 700 frames for fibrinogen and IgG1 and 11200 frames for BSA. Several well-defined diffraction-limited iSCAT spots in Fig. 2a-c reveal typical contrasts ranging from $3\times10^{-4}$ for BSA to $1\times10^{-3}$ for fibrinogen. The standard deviation of the residual background fluctuations in the differential images amounts to $2.5\times10^{-4}$ for averages of 700 frames (i.e. for fibrinogen and IgG1) and $9\times10^{-5}$ for 11200 frames (BSA). We have verified that the contrast scales as the square root of the integration time, confirming that our detection has reached the shot noise limit. A quantitative account of the signal-to-noise is presented in the Supplementary Information.

Since the iSCAT signal of all individual proteins of a certain type should be the same within our flat field of view, we expect a narrow distribution of the observed contrasts as long as the arrival rate of the analyte is lower than the acquisition rate. To examine our data in this respect, we searched the recorded differential images for local minima that appeared over an area of one point-spread function. The top row in Fig. 2a-c displays the histograms of the magnitudes of the iSCAT minima obtained prior (red) and during (black) the injection of the analyte. In each histogram, a clear black peak distinguishes the signal from the background contribution. The fact that the width of the signal distribution is always much smaller than its peak value and that each distribution rapidly falls at larger contrasts lets us attribute each peak to the contrast of a single protein. The small number of occurrences at the large tails of the distributions indicates rare events of clusters of two or more proteins.

Fig. 2d plots the relationship between the measured iSCAT contrast and the molecular masses of the three proteins. Here, we expect a linear relation because the iSCAT signal is proportional to the protein polarizability and therefore its volume if we assume similar indices of refraction and densities for different proteins. Experimental confirmation of this prediction in Fig. 2d provides further evidence for the robustness of our measurements and interpretation.

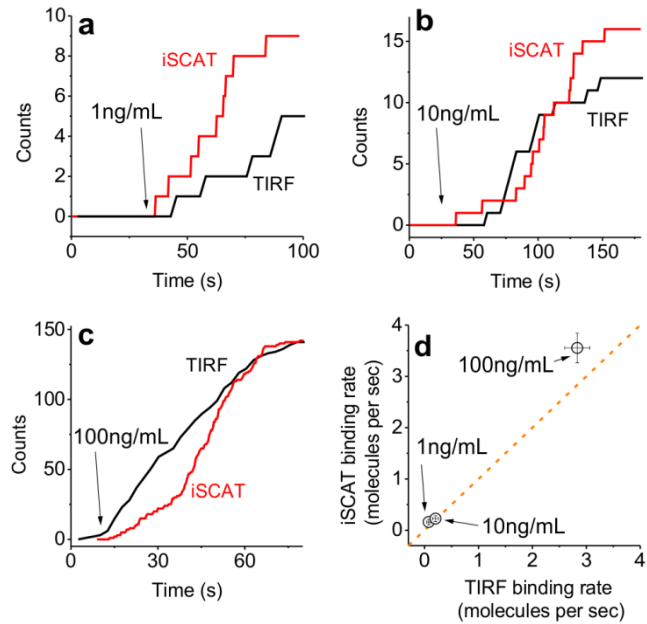

**Fig. 3 IgG binding at different concentrations. a, b, c,** Kinetics of iSCAT and fluorescence signals (measured consecutively) for IgG solutions with concentrations 1 ng/mL, 10ng/mL, and 100 nm/mL, respectively. **d**, Plot of the iSCAT binding rate versus the binding rate determined by TIRF. The diagonal line marks a slope of one. Error bars indicate the statistical error resulting from the total number of detected molecules.

To demonstrate the practical biosensing capabilities of iSCAT, we functionalized a coverslip with anti-IgG1 antibodies and used it to detect different concentrations of IgG1. For a control experiment, IgG1 was labeled with Alexa 647 dye and detected with single-molecule sensitivity via total internal reflection fluorescence (TIRF) microscopy. Here, we carried out the fluorescence detection on the same sensor surface while the iSCAT illumination was temporarily switched off to avoid strong photobleaching. Series of 1ng/mL, 10ng/mL, and 100ng/mL concentrations of IgG1+Alexa647 in phosphate buffered saline (PBS) were pumped to the detection area through the micropipette for about one minute. TIRF and iSCAT images were recorded subsequently. The red and black curves in Fig. 3a-c display time series of individual binding events for iSCAT and TIRF, respectively. In all cases, we find a stable baseline prior to the detection followed by a rapid increase in the binding rate, ranging from 5 to 150 bindings per minute for different concentrations. Fig. 3d shows a linear relation between the binding rates measured by iSCAT and TIRF. Considering the single-molecule sensitivity of TIRF detection, this outcome provides another strong independent evidence that our label-free iSCAT biosensor has reached the single-molecule detection level. Moreover, the data show that an amount of analyte as small as 3 attomoles, corresponding to a concentration of 1ng/mL was sufficient for achieving a good SNR.

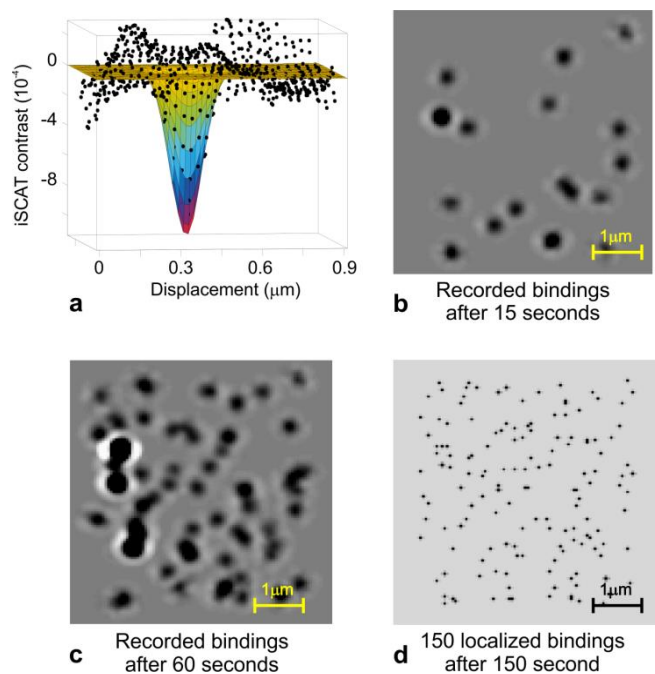

**Fig. 4 Superresolution imaging. a**, An iSCAT image of a molecule fitted with a two-dimensional Gaussian, yielding a localization precision of 5 nm. **b**, **c**, Images of individual molecules accumulated in 15 s and 60 s, respectively. **d**, Superresolution image, showing the localized positions of the binding events accumulated in 150 s.

In addition to the ultimate biosensor performance of detecting individual analyte molecules, our recordings register the spatial coordinates of each molecule with nanometer precision. The symbols in Fig. 4a represent the iSCAT image of a single fibrinogen while the surface plot depicts a two-dimensional Gaussian fit. The SNR of the order of 10 allows localization of the center of the Gaussian peak with a precision of 5 nm, which agrees with the theoretical limit of localization within 10%[25]. Since in our experiment the proteins land one by one, we can extend this procedure to acquire supper-resolution iSCAT images of the binding process, whereby subsequent arrival times are used to identify the signals of individual molecules[26]. Fig. 4b and c show iSCAT images accumulated after 15 s and 60 s of detection, and Fig. 4d displays the localization map of all molecules accumulated in 150 s. This provides the first generalization of the recent super-resolution microscopy methods[27,28] to nonfluorescent samples.

We have shown that contrary to the common wisdom, careful consideration of the quantities involved and proper experimental procedure make it is possible to detect the Rayleigh scattering of a single unlabeled biomolecule in a simple and direct optical measurement. This approach is not limited to confined optical fields, can count proteins, is compatible with a wide range of functionalization methods, provides nanoscopic spatial information of binding events, and can be easily parallelized. In addition, iSCAT sensing can be used to visualize and monitor the association and dissociation kinetics of biomolecules and study their cooperative interactions[29] because it does not suffer from photobleaching. The sensitivity of iSCAT in our experiment was determined by the pixel well depth of the camera. Considering that the fundamental limit of this method is set by photon shot noise, it can be improved at higher incident powers or integration times. This would allow one to detect yet smaller biomolecules such as microRNA[30] or environmental pollutants. Higher signal-to-noise ratios might also open the door to extracting the absolute molecular mass, shape and orientation of single molecules via the measurement of their polarizabilities. Together with its ease of instrumentation, these features make iSCAT sensing highly promising for multiplexed sensor arrays in laboratory and industrial applications alike.

**Acknowledgements:**

This work was supported by the Max Planck Society and an Alexander-von-Humboldt Professorship.

[*] To whom correspondence should be addressed: Email: vahid.sandoghdar@mpl.mpg.de

**Materials**

Poly-ethyleneglycol-coated coverslips with N-hydroxysuccinimide (NHS) coupling chemistry for covalent immobilization of proteins were purchased from MicroSurfaces, Inc, TX, USA. Fibrinogen from human plasma, mouse-IgG1 (whole antibody), goat-anti-mouse-IgG1 (whole antibody), mouse-IgG1 labeled with Alexa Fluor® 647, and Ultrapure Bovine Serum Albumin (BSA) were purchased from Life Technologies GmbH, Germany. Phosphate Buffered Saline (PBS), Sodium Acetate buffer (SA, pH 5.2) and other common chemical were obtained from Sigma-Aldrich.

**Methods**

The output of a continuous-wave diode laser (wavelength 405, Toptica, Inc.) was intensity stabilized to better than $3\times10^{-4}$ using a proportional-integral-derivative control loop. The incident beam was circularly polarized and directed through a 70:30 beam-splitter and focused at the back focal plane of a microscope objective (NA=1.46, Zeiss). Functionalized surface of a glass coverslip was placed in the focus of the microscope objective and the focal position was stabilized with an active feedback loop. The full width at half-maximum (FWHM) of the microscope point-spread-function (PSF) was typically between 190 nm – 200 nm. A narrow field of view of 4.5 µm × 4.5 µm was illuminated with a power of 10mW with field flatness variations below 1%. The light reflected and scattered at the glass/water interface was collected with the same microscope objective and imaged via the 70:30 beam splitter on a fast CMOS camera (Photonfocus, AG). An area of 128x128 pixels of the camera was acquired at the frame rate of 3000 frames per second.

In order to visualize changes in the scattering signal and their spatial distribution, we continuously processed sets of several hundreds of consecutive frames and extracted the image of the surface roughness (see Fig. 1b,c). Differential images were then calculated from two sets of images separated by a temporal delay of 300 ms and a running average was used to accomplish continuous acquisition of differential images. Any mechanical drifts and instabilities were negligible on the time scale of up to 5 s. The SNR of the differential images was high enough to resolve changes at the level of 1% of the surface roughness. The standard deviation of the differential images amounted to $2.5\times10^{-4}$ for an average of 700 frames. The noise decreased with the square root of the frame number up to about 25000 frames, where other system instabilities began to prevail.

A pure diluting buffer was placed in a plexiglass cuvette of 5 mL volume on the coverslip. Micropipettes were pulled from a thin-wall capillary (OD 1 mm / ID 0.75 mm) to obtain flat ends with an opening diameter of 5 μm. The micropipette was positioned at about 10 μm above the coverslip surface and in close proximity to the iSCAT field of view. This position was chosen to avoid visible artifacts or fringes in the image. The characteristic volume between the pipette tip and the surface was in the picoliter range. This volume results in a diffusion time of about 200 ms for the analyte molecules to reach the sensor surface. The minimum flow rate required to maintain constant concentration in the vicinity of the sensor surface was of the order of nL/min. Higher flow rates of 500 nL/min driven by a conventional syringe pump were used to maintain a stable pulse-free flow. Coverslips were coated with poly(ethyleneglycol) (PEG) brush (thickness of 2-3 nm) with NHS reactive groups ($10^7$ binding sites within the field of view). Coated coverslips were either used directly to capture BSA, IgG1 or fibrinogen or further functionalized with high concentration of anti-IgG1 (20 μg/mL). In the first case, the cuvette was filled with a buffer of sodium acetate (SA). 200ng/mL solution of the protein in SA buffer was loaded in the micropipette. After acquiring 100 frames of the baseline signal the sample was pumped for a few minutes in order to clearly observe the difference of the surface scattering before and during the detection. Once the pumping was stopped, the analyte molecules diffused and diluted in the cuvette on the time scale of seconds, decreasing the concentration at the surface by typically 3 orders of magnitude (depending on the duration of pumping). In the case of specific IgG1 detection, the sensor surface was incubated with 20 μg/mL solution of anti-IgG1 in SA buffer for 20 minutes followed by 20 minutes incubation in NHS deactivating buffer. The functionalized coverslip was then mounted on the setup, the cuvette was filled with PBS buffer, and the micropipette with the corresponding target concentration (1 ng/mL, 10 ng/mL, or 100 ng/mL) of IgG1 dissolved in PBS was placed into the cuvette.

# Supplementary Information for

# Direct optical sensing of single unlabeled small proteins and super-resolution microscopy of their binding sites

Marek Piliarik and Vahid Sandoghdar

*Max Planck Institute for the Science of Light and Friedrich Alexander University Erlangen-Nuremberg*

## A- Detection and data processing

Considering that the iSCAT signal from a single protein is much smaller than the iSCAT contribution from the sensor surface, a key step is to account for the latter. To do this, we continuously processed sets of several hundreds of consecutive frames and determined the image of the surface roughness (see Figure 1S). Differential images were then calculated from two sets of images separated by a temporal delay of 300 ms and a running average was used to perform continuous acquisition of differential images.

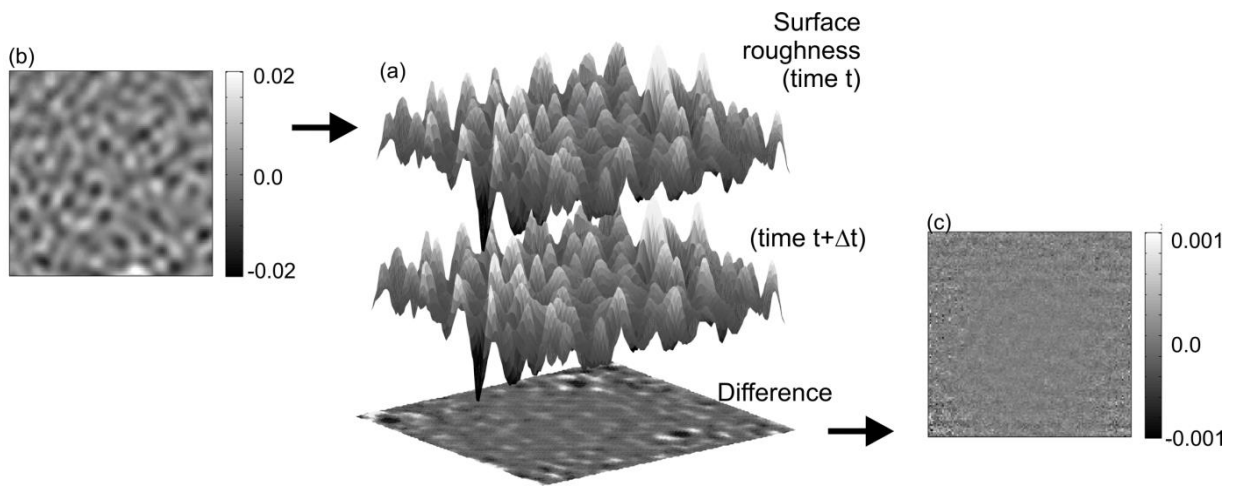

**Figure 1S** (a) The principle of background subtraction. Two images of surface roughness (b) as measured on the camera with a temporal offset of $\Delta t$ = 300 ms are subtracted, and the resulting image of the temporal changes on the surface in (c) shows the shot-noise-limited background fluctuations.

Figure 2S shows time series of differential images acquired during the detection of fibrinogen. The first three differential images show the background signal before introducing the analyte. When the sample is pumped into the proximity of the sensor surface clear and discrete spots of molecular bindings appear in the image (five examples shown in Figure 2S). After the sample flow is stopped, the differential images show only the background noise again.

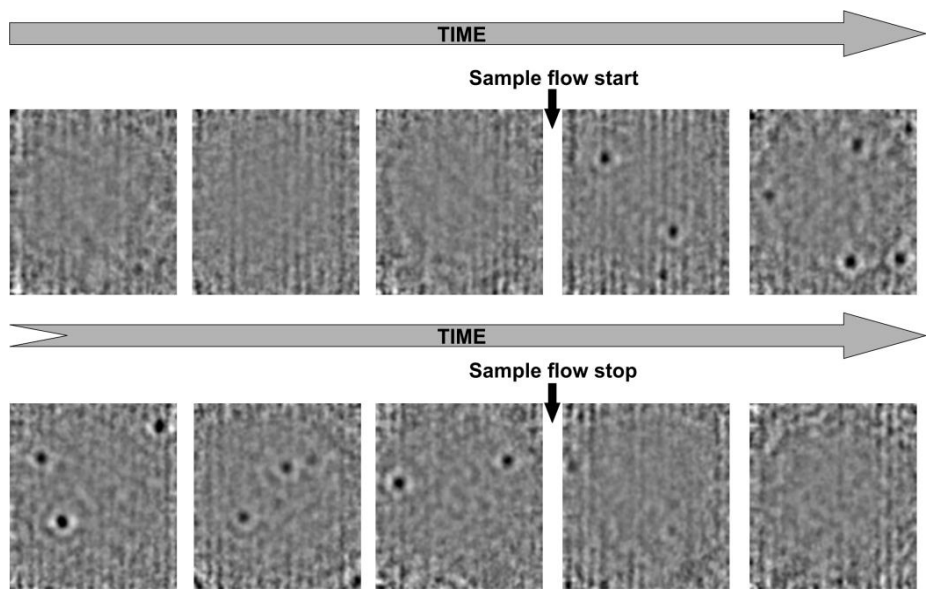

**Figure 2S** Time series of differential images acquired during the detection of fibrinogen.

All differential images were processed with a routine that searched for local minima over a region of the size of a diffraction-limited spot on the camera (CDLS). The histograms of the data obtained during the detection of each molecular sample as well as prior to the injection of the sample were calculated and are discussed in the main paper (see Figure 2 (a-c)). The background histograms correspond to the case where no molecules were bound to the surface and indicate that the minimum-finding algorithm also identifies local minima originating in the background noise.

## B- Single molecule visibility and detector noise

As discussed in Lindfors et al, PRL **93**, 037401 (2004), the absolute contrast of iSCAT and thus the total number of photons on the detector can be adjusted through the reflectivity *r*. However, if the measurement is shot-noise limited, the dependence of the visibility of the iSCAT signal on *r* drops out because both the shot noise and the interference term are linearly proportional to *r*.

Below, we present a quantitative estimate of our experimental parameters for the detection of the smallest protein (BSA), leading to a noise and visibility analysis.

- We illuminate 20 µm² of the sensor surface with 10 mW incident power, which amounts to 15 µW per the area of diffraction-limited spot (DLS).

- Considering the photon energy at the wavelength 405 nm of 5×10⁻¹⁹J,
$$P_{inc} = 3 \times 10^{13} \text{photons s}^{-1}\text{DLS}^{-1}.$$

- For a scattering cross section of $\sigma = 10^{-11}$ µm², collection efficiency of 35%, losses of 75% in the detection path (i.e. 0.25 throughput), and area of DLS $= \pi(100 \, nm)^2$ in our experimental setup, the detector receives
$$P_{scat} = \frac{10^{-11} \, \mu m^2}{\pi(0.1 \, \mu m)^2} \times 0.35 \times 0.25 \, P_{inc} = (3 \times 10^{-11}) P_{inc}.$$

- Assuming $r^2 = 0.006$ for the reflectivity of glass-water interface, the powers in reference beam and the interferometric term become:

$$P_{ref} = 6 \times 10^{-3} \times 0.25 \, P_{inc} = 1.5 \times 10^{-3} P_{inc} = 4.5 \times 10^{10} \text{photons s}^{-1}\text{DLS}^{-1},$$

$$P_{int} = 2\sqrt{P_{scat} P_{ref}} = 2 \times \sqrt{(3 \times 10^{-11}) \times (1.5 \times 10^{-3})} \, P_{inc} = 4 \times 10^{-7} P_{inc}$$
$$= 2.7 \times 10^{-4} P_{ref} = 10^7 \text{photons s}^{-1}\text{DLS}^{-1}.$$

- Our experimental setup is configured to provide 300x magnification with a camera pixel size of 10µm. Therefore, the area of a diffraction-limited spot on the camera (CDLS) corresponds to:

$$\text{CDLS} = \frac{\pi(300 \times 0.1 \mu m)^2}{(10 \mu m)^2} = 30 \text{ camera pixels.}$$

- Considering a quantum efficiency *QE*=0.25 for the CMOS camera at wavelength 405 nm and exposure time of *t* = 0.2 ms, the reference and interferometric signals received by each pixel are:

$$\frac{P_{ref}}{\text{pixel}} = QE \frac{P_{ref} \times t}{\text{CDLS}} = \frac{0.25 \times (4.5 \times 10^{10}) \times (2 \times 10^{-4})}{30}$$
$$= 8 \times 10^4 \text{ photoelectrons,}$$

The well depth of the CMOS camera was $10^5$ photoelectrons, limiting the maximum incident power in our experiment.

In a similar fashion, we can calculated the contribution of the iSCAT signal:

$$\frac{P_{int}}{\text{pixel}} = QE \frac{P_{int} \times t}{\text{CDLS}} = \frac{0.25 \times (10^7) \times (2 \times 10^{-4})}{30} = 16 \text{ photoelectrons}.$$

- To achieve a better shot-noise limit, we averaged many frames.

$$11200 \, P_{int}/\text{pixel} = 2 \times 10^5,$$

while the standard deviation of the signal fluctuation becomes

$$\sigma_{11200 pixels} = \sqrt{11200 \, P_{ref}/pixel} = \sqrt{11200 \times 8 \times 10^4} = 3 \times 10^4 \text{ photoelectrons}.$$

*In this case, we obtain $\frac{P_{int}}{\sigma} \cong 10$, which is in good agreement with our experimental results.*

To show that our detection reaches the shot noise limit, we have analyzed images recorded from the sensor surface (without any analyte) by comparing the results of different averaging processes. Figure 3S (left) plots the histograms of the amplitudes of iSCAT minima calculated using 700, 2800 and 11200 frames for averaging. The dependence of the minimum distribution and image standard deviation (SD) on frame averaging is presented in Figure 3S (right). We find that the detection is very close to the theoretical shot-noise limit. At very long integration times, drifts and instabilities of the setup begin to dominate shot noise.

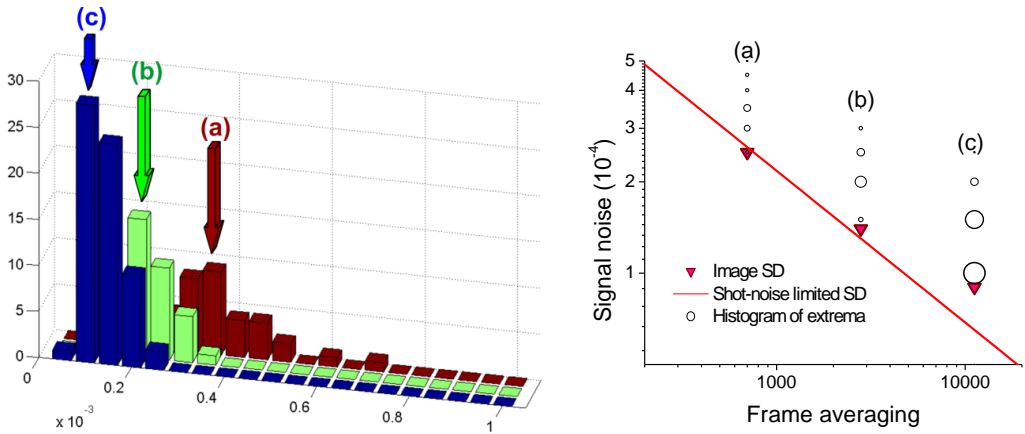

**Figure 3S** (left) Comparison of histograms of background noise calculated for (a) 700-frame running average (b) 2800-frame running average, and (c) 11200-frame running average. (Right) The plot shows the dependance of the histograms on the frame averaging in a buble graph (circles). Standard deviation of the images is plotted by triangles, and the solid line indicates a theoretical limit of the shot noise determined by the square root of the number of frames.